\documentclass[12pt,showpacs]{revtex4-1}%

\usepackage[english]{babel}
\usepackage[T1]{fontenc}
\usepackage{bbm}
\usepackage{epsfig}
\usepackage{psfrag}
\usepackage{color}
\usepackage{amsmath}
\usepackage{amsfonts}
\usepackage{amssymb}
\usepackage{graphicx}%
\newcommand{\be}{\begin{equation}}
\newcommand{\ee}{\end{equation}}
\newcommand{\bea}{\begin{eqnarray}}
\newcommand{\eea}{\end{eqnarray}}

\newcommand{\Q}{{\bf a}}

\renewcommand{\theequation}{\arabic{section}.\arabic{equation}}

\begin{document}

\title{Exact self-accelerating cosmologies  in the ghost-free 
massive gravity -- the detailed derivation
}

\author{Mikhail~S.~Volkov}

\affiliation{
Laboratoire de Math\'{e}matiques et Physique Th\'{e}orique CNRS-UMR 7350, 
Universit\'{e} de Tours, Parc de Grandmont, 37200 Tours, FRANCE}

\email{volkov@lmpt.univ-tours.fr}

\begin{abstract}

We present the detailed derivation of the recently announced 
most general cosmological solution with homogeneous and isotropic metric in the 
 ghost-free massive gravity theory. 
We use the standard parametrization of the theory in terms
of the matrix square root, and then show how the same results are recovered within the 
tetrad formulation.  
The solution obtained includes the matter source, it exists 
for generic values of the theory parameters, and it describes a universe that can be 
spatially open, closed, or flat, and that shows the late time acceleration due to the 
effective cosmological term mimicked by the graviton mass. The  St\"uckelberg fields
are inhomogeneous, which could probably give rise to inhomogeneous perturbations
of the homogeneous and isotropic backgrounds, although
this effect should be suppressed by the smallness of the graviton mass.

\end{abstract}

\maketitle

\section{Introduction} 

Considering theories with massive gravitons \cite{Pauli} 
is motivated 
by the observation of the current acceleration of our universe \cite{Reiss}, 
since the graviton mass can induce an effective cosmological term. 
However, 
for a long time 
such theories were not considered as being suitable  
for describing the real world, since they typically
contain the Boulware-Deser ghost \cite{BD}. This is an unphysical mode 
with negative norm that propagates together with the physical 
degrees of freedom  and renders unstable generic curved
 backgrounds in the theory. 
Fortunately, it was recently discovered that the presence of the ghost 
is not mandatory, since there exist a special  
massive gravity theory   
which is ghost-free \cite{RGT}. A careful analysis shows 
that the number of propagating degrees of freedom in this theory 
agrees with the number of graviton polarizations,
so that there are no extra unphysical propagating modes  \cite{noghost}. 
This does not mean that all backgrounds are stable in this theory,
since there could be other instabilities, which should be checked in each 
particular case.  However, since 
the most dangerous instability is absent, 
the ghost-free theory of massive gravity can be 
considered as a healthy  physical model that can be used 
for interpreting the observational data. This motivates studying cosmological
solutions in this theory.

The first self-accelerating cosmologies  in the ghost-free massive gravity 
were obtained without matter, they describe the pure de Sitter universe \cite{Koyama}. 
The matter source was then included for special values of the theory 
parameters \cite{cosm}. A surprising feature of these solutions is that their physical 
and reference  metrics do not share the same Killing symmetries -- although 
the physical metric is of the Friedmann-Robertson-Walker (FRW) type,   
the reference  metric is inhomogeneous. This creates a technical difficulty, 
since the equations for the 
St\"uckelberg fields reduce to a non-linear partial differential equation (PDE). 

One more similar solution was found 
in Ref.\cite{gaba}, where it was argued that, even though 
the background geometry is homogeneous and isotropic, the inhomogeneous structure of the 
St\"uckelberg fields should be visible when the background is perturbed. 
However, this effect should be strongly suppressed by the smallness of the graviton mass,
so that in small compared to the graviton Compton wavelength regions the deviations 
from the standard FRW cosmology should be small. These properties seem to 
be generic for massive gravity cosmologies, since, as was noticed in \cite{gaba},
the `genuinely'  homogeneous and isotropic solutions for which both metrics would be FRW
do not exist in the theory, at least when the metrics are spatially flat. 
Even though such solutions were later found in the spatially open case, they are 
less interesting physically, and show in addition a nonlinear instability \cite{Muk}.

It seems therefore that physical cosmologies with massive gravitons 
should be described by solutions with a homogeneous and isotropic physical metric,
but with an inhomogeneous or anisotropic (or both) reference   metric
(unless the latter is chosen to be non-flat \cite{DS}). 
Such solutions were also constructed \cite{cosm1} in the ghost-free bigravity \cite{HR1}, 
when both metrics are dynamical but are not simultaneously diagonal and do not share 
the same symmetries. However, an all cases the solutions were obtained
only for constrained and not generic values of the theory parameters. 
Very recently, a solution for generic parameter values 
was announced in \cite{Wyman}, but without determining the St\"uckelberg fields,
so that the most difficult part of the problem was actually skipped. 
Finally, the complete solution was obtained in \cite{vvv}, both within the bigravity
and massive gravity. In what follows we shall present a detailed derivation of this result. 

For pedagogical reasons, we shall restrict our discussion below only 
to the massive gravity case, since 
in the bigravity the procedure is essentially the same but the formulas are more complicated. 
The analysis of \cite{vvv} was carried out within the tetrad formalism, but 
we shall employ  below the standard parametrization 
in terms of the matrix square root used by most authors. 
At the same time, we shall show how the same results are recovered within the tetrad formulation.  
We shall try to be maximally explicit. For example, it turns out that it is not easy to find 
in the literature the explicit form of the field equations in the theory, 
and most authors prefer to put a symmetry ansatz to the action and then vary. 
We shall therefore show how to vary the action in the general case.     
We shall also show that the tetrad formulation  is equivalent to the standard one. 

We make the ansatz for which the physical metric is FRW but the reference  metric 
is only spherically symmetric and not diagonal. We then calculate the matrix square root
and show that the resulting equations are such that the St\"uckelberg scalars 
effectively decouple, and the metric satisfies Einstein equations 
containing on the right an effective 
cosmological term and the matter source. 
However, such a decoupling is only possible if the scalars satisfy a 
consistency condition expressed by a complicated non-linear PDE. Fortunately, the latter
can be solved exactly.   
As a result, we obtain the most general cosmological solution 
for which the physical metric is homogeneous and isotropic 
but the St\"uckelberg fields are inhomogeneous.    
The solution includes the matter source, it exists for generic parameter values, 
and it describes a FRW universe 
that can be spatially flat, open or closed, and which shows 
the late-time acceleration due to the effective cosmological term mimicked
by the graviton mass. 

The rest of the paper is organized as follows. 
In Sections II, III we introduce the ghost-free massive gravity and derive its 
equations of motion. The symmetry reduction is described in Section IV. 
The solution for the metric is presented in Section V, while the 
consistency condition for its existence is analyzed in 
Section VI. In Section VII all results are rederived again within 
the tetrad formulation. 
Finally, in the Appendix we describe all possible 
solutions for which both metrics are homogeneous and isotropic.

\section{The ghost-free massive gravity}
\setcounter{equation}{0}

The ghost-free massive gravity theory 
of de Rham, Gabadadze, and Tolley (dRGT) 
\cite{RGT} is defined on a four-dimensional 
spacetime manifold equipped with the metric $g_{\mu\nu}$ and carrying 
four scalar fields $X^A$ (St\"uckelberg scalars) which parameterize 
the reference  metric 
$
f_{\mu\nu}=\eta_{AB}\partial_\mu X^A\partial_\nu X^B
$
with $\eta_{AB}={\rm diag}[1,-1,-1,-1]$.
The action is 
\bea                                  \label{action}
S&=&\frac{1}{8\pi G}\,\int \left(-\frac{1}{2}\,R
+m^2{\cal U}\right)
\sqrt{-g}\,d^4x
+S_{\rm m}\,,
\eea
where 
$m$ is the graviton mass and $S_{\rm m}$ describes the ordinary matter
(as for example perfect fluid) that interacts with $g_{\mu\nu}$ in the usual way.  
To define ${\cal U}$, 
the interaction between $g_{\mu\nu}$ and $f_{\mu\nu}$, 
the key element is the tensor $$\gamma^\mu_{~\nu}=\sqrt{g^{\mu\alpha}f_{\alpha\nu}},$$
where $g^{\mu\nu}$ is the inverse of $g_{\mu\nu}$. Here the square root
is understood in the sense that 
\be                     \label{gam}
(\gamma^2)^\mu_{~\nu}\equiv \gamma^\mu_{~\alpha}\gamma^\alpha_{~\nu}=g^{\mu\alpha}
f_{\alpha\nu},
\ee 
or, using the hat to denote matrices, 
\be
\hat{\gamma}^2=\hat{g}^{-1}\hat{f}.
\ee
Introducing ${\cal K}^\mu_\nu=\delta^\mu_\nu-\gamma^\mu_{~\nu}$ with 
traces $[{\cal K}]\equiv {\rm tr}(\hat{{\cal K} })={\cal K}^\mu_\mu$
and $[{\cal K}^n]\equiv {\rm tr}(\hat{{\cal K} }^n)=({\cal K}^n)^\mu_\mu$,
the interaction is
\be                                \label{int}
{\cal U}={\cal U}_2+\alpha_3\,{\cal U}_3+\alpha_4\,{\cal U}_4\,,
\ee
where $\alpha_3,\alpha_4$ are parameters and
%\begin{subequations}
\begin{align}      \label{U1}
{\cal U}_2&=\frac{1}{2!}\,([{\cal K}]^2-[{\cal K}^2]),\notag \\
{\cal U}_3&=\frac{1}{3!}\,([{\cal K}]^3-3[{\cal K}][{\cal K}^2]+2[{\cal K}^3]),\notag \\
{\cal U}_4&=\frac{1}{4!}\,([{\cal K}]^4-6[{\cal K}]^2[{\cal K}^2]
+8[{\cal K}^3][{\cal K}]+3[{\cal K}^2]^2-6[{\cal K}^4]).
\end{align}
%\end{subequations}
Equivalently, with $\lambda_A$ being eigenvalues of $\hat{{\cal K} }$,
\begin{align}     \label{U2}
{\cal U}_2=\sum_{A<B}\lambda_A\lambda_B,~~~~~
%\lambda_1\lambda_2+\lambda_1\lambda_3+
%\lambda_1\lambda_4+\lambda_2\lambda_3+\lambda_2\lambda_4+
%\lambda_3\lambda_4,\notag  \\
{\cal U}_3=
\sum_{A<B<C}\lambda_A\lambda_B\lambda_C,~~~~~
%=\lambda_1\lambda_2\lambda_3+\lambda_1\lambda_2\lambda_4+
%\lambda_1\lambda_3\lambda_4+\lambda_2\lambda_3\lambda_4,\notag \\
{\cal U}_4=%\sum_{A<B<C<D}\lambda_A\lambda_B\lambda_C\lambda_D
\lambda_1\lambda_2\lambda_3\lambda_4={\rm det}(\hat{{\cal K} }). 
\end{align}
One more equivalent representation is (with $\epsilon^{0123}=-\epsilon_{0123}=1$) \cite{Niew} 
%\begin{subequations}
\begin{align}      \label{U3}
{\cal U}_2&=-\frac{1}{2!}\,\epsilon_{\mu\nu\rho\sigma}
\epsilon^{\alpha\beta\rho\sigma}{\cal K}_{\alpha}^{\mu}
{\cal K}_{\beta}^{\nu} ,\notag \\
{\cal U}_3&=-\frac{1}{3!}\,\epsilon_{\mu\nu\rho\sigma}
\epsilon^{\alpha\beta\gamma\sigma}{\cal K}_{\alpha}^{\mu}
{\cal K}_{\beta}^{\nu}{\cal K}_{\gamma}^{\rho},\notag \\
{\cal U}_4&=-\frac{1}{4!}\,\epsilon_{\mu\nu\rho\sigma}
\epsilon^{\alpha\beta\gamma\delta}{\cal K}_{\alpha}^{\mu}
{\cal K}_{\beta}^{\nu}{\cal K}_{\gamma}^{\rho}{\cal K}_{\delta}^{\sigma}\,.
\end{align}
%\end{subequations}

\section{Field equations}
\setcounter{equation}{0}

Some care is needed when varying the action with respect to the metric. 
Let us first vary the constraint  \eqref{gam},  
\be
\delta\hat{\gamma}\hat{\gamma}+\hat{\gamma}\delta\hat{\gamma}=\delta \hat{g}^{-1}\hat{f},
\ee
which gives
\be                              \label{del}
\delta\hat{\gamma}+\hat{\gamma}\delta\hat{\gamma}\hat{\gamma}^{-1}
=\delta \hat{g}^{-1}\hat{f}\hat{\gamma}^{-1}.
\ee
This cannot be resolved with respect to $\delta\hat{\gamma}$, because 
$\hat{\gamma}$ and  $\delta\hat{\gamma}$ do not commute. However,
only the matrix traces enter the action.
Taking the trace of \eqref{del} gives 
\be                              \label{del1}
{\rm tr}(\delta\hat{\gamma})+{\rm tr}( \hat{\gamma}\delta\hat{\gamma}\hat{\gamma}^{-1})=
2{\rm tr}(\delta\hat{\gamma})
={\rm tr}(\delta \hat{g}^{-1}\hat{f}\hat{\gamma}^{-1}),
\ee  
and noting that $\hat{f}=\hat{g}\hat{g}^{-1}\hat{f}=\hat{g}\hat{\gamma}^2$\,,
one obtains
\be                              \label{del2}
\delta [{\cal K}]=-{\rm tr}(\delta\hat{\gamma})
=-\frac12\,{\rm tr}(\delta \hat{g}^{-1}\hat{g}\hat{\gamma}).
\ee  
Next, one has 
\be                              \label{del3}
\delta [{\cal K}^{n+1}]=\delta\,{\rm tr}(\hat{\cal K}^{n+1})=(n+1){\rm tr}(\delta\hat{\cal K}\,
\hat{\cal K}^{n})=-(n+1){\rm tr}(\delta\hat{\gamma}\,
\hat{\cal K}^{n}),
\ee 
where the cyclic property of the trace is used. Now, 
multiplying \eqref{del} by $\hat{\cal K}^{n}$
from the right,
taking the trace and using the fact that $\hat{\gamma}$ and $\hat{\cal K}$ commute, 
one obtains 
\be                              \label{del4}
2{\rm tr}(\delta\hat{\gamma}\,\hat{\cal K}^{n})
={\rm tr}(\delta \hat{g}^{-1}\hat{f}\hat{\gamma}^{-1}\,\hat{\cal K}^{n}),
\ee  
which finally gives 
\be                              \label{del5a}
\delta [{\cal K}^{n+1}]
=-\frac{n+1}{2}\,{\rm tr}(\delta \hat{g}^{-1}\hat{g}\hat{\gamma}\hat{\cal K}^{n})=
-\frac{n+1}{2}\,\delta g^{\mu\nu}g_{\nu\beta}\gamma^\beta_{~\sigma}
({\cal K}^{n})^\sigma_\mu\,.
\ee  
It turns out that the matrix 
$\hat{g}\hat{\gamma}\hat{\cal K}^{n}$
is actually symmetric \cite{Visser}. 
To see this, one uses the relation ($m$ is a non-negative integer)
\be                            \label{rel1}
\hat{g}\left(\sqrt{\hat{g}^{-1}\hat{f}}  \right)^m\hat{g}^{-1}=
\left(\sqrt{\hat{f}\hat{g}^{-1}}  \right)^m,
\ee
which follows from the fact that squaring it gives identity.  
This implies that 
\be                          \label{rel2}
\hat{g}\left(\sqrt{\hat{g}^{-1}\hat{f}}  \right)^m=
\left(\sqrt{\hat{f}\hat{g}^{-1}}  \right)^m\hat{g}\,,
\ee
and hence
\be                          \label{rel2a}
\left(\hat{g}\left(\sqrt{\hat{g}^{-1}\hat{f}}  \right)^m\right)^{\rm tr}=
\left(\left(\sqrt{\hat{f}\hat{g}^{-1}}  \right)^m\hat{g}\right)^{\rm tr}=
(\hat{g})^{\rm tr}\left(\sqrt{(\hat{g}^{-1})^{\rm tr}
(\hat{f})^{\rm tr}}  \right)^m=
\hat{g}\left(\sqrt{\hat{g}^{-1}\hat{f}}  \right)^m,
\ee
so that $\hat{g}\hat{\gamma}^m$ is a symmetric matrix. Since 
$\hat{g}\hat{\gamma}\hat{\cal K}^{n}$ can be represented as the 
sum of terms of the form $\hat{g}\hat{\gamma}^m$
with different $m$, it is also symmetric. This finally gives 
\be                              \label{del5}
\frac{\delta [{\cal K}^{n+1}]}{\delta g^{\mu\nu}}
=-\frac{n+1}{2}\,g_{\nu\beta}\gamma^\beta_{~\sigma}
({\cal K}^{n})^\sigma_\mu\,,
\ee  
and the expression on the right here is symmetric with respect to $\mu\leftrightarrow \nu$.

It is now straightforward to vary the action. 
This gives the Einstein equations 
\be
G_{\mu\nu}=m^2T_{\mu\nu}+8\pi G T^{\rm (m)}_{\mu\nu}\,,
\ee
where $T^{\rm (m)}_{\mu\nu}$ is the matter energy-momentum tensor obtained by varying
$S_{\rm m}$, while 
\be                              \label{TT1}
T_{\mu\nu}=2\frac{\delta {\cal U}}{\delta g^{\mu\nu}} -{\cal U}\,g_{\mu\nu}\,.
\ee
Using ${\cal U}$ in \eqref{int},\eqref{U1} and applying 
\eqref{del5} gives 
\bea                                \label{TT3}
T_{\mu\nu}&=&
\gamma_{\mu\alpha}\left\{{\cal K}^\alpha_\nu-[{\cal K}]\delta^\alpha_\nu\right\}\notag \\
&-&\alpha_3\gamma_{\mu\alpha}\left\{{\cal U}_2\,\delta^\alpha_\nu
-[{\cal K}]{\cal K}^\alpha_\nu +({\cal K}^2)^\alpha_\nu \right\} \notag \\
&-&\alpha_4\gamma_{\mu\alpha}\left\{{\cal U}_3\,\delta^\alpha_\nu
-{\cal U}_2\,{\cal K}^\alpha_\nu
+[{\cal K}]({\cal K}^2)^\alpha_\nu -({\cal K}^3)^\alpha_\nu
\right\}\notag \\
&-&{\cal U}\,g_{\mu\nu}\,,
\eea
where $\gamma_{\mu\alpha}=g_{\mu\sigma}\gamma^\sigma_{~\alpha}$. 
The above considerations guarantee that $T_{\mu\nu}=T_{(\mu\nu)}$. 
It is also worth mentioning the equivalent representation, 
\bea                              \label{tau1}
T_{\mu\nu}=\gamma_{\mu\alpha}\left\{{\cal K}^\alpha_\nu-[{\cal K}]\delta^\alpha_\nu
+\alpha_3 {\Xi}^\alpha_\nu+\alpha_4 {\Omega}^\alpha_\nu
\right\}-{\cal U}\,g_{\mu\nu}\,,
\eea 
where 
\bea                                  \label{tau2}
{\Xi}^\alpha_\nu&=&\frac{1}{2!}\,\epsilon_{\nu\mu\rho\sigma}
\epsilon^{\alpha\beta\gamma\sigma}{\cal K}_{\beta}^{\mu}{\cal K}^{\rho}_{\gamma},\notag \\
{\Omega}^\alpha_\nu&=&\frac{1}{3!}\,\epsilon_{\nu\mu\rho\sigma}
\epsilon^{\alpha\beta\gamma\delta}{\cal K}_{\beta}^{\mu}
{\cal K}^{\rho}_{\gamma}{\cal K}^{\sigma}_{\delta}.
\eea
Since the matter energy-momentum tensor is conserved 
due to the diffeomorphism-invariance of the matter action $S_m$, 
\be                         \label{matter0}
\nabla^\mu T^{(m)}_{\mu\nu}=0,
\ee
the Bianchi identities for the Einstein equations 
imply the conservation condition,
\be                      \label{Stuck}
\nabla^\mu T_{\mu\nu}=0,
\ee
which can be viewed as equations for the 
St\"uckelberg fields. 

\section{Symmetry reduction and taking the square root \label{sym} }
\setcounter{equation}{0}

Let us now choose spherical coordinates $x^\mu=(t,r,\vartheta,\varphi)$
and assume the physical metric to be homogeneous and isotropic,
\be                         \label{gg}
ds_g^2=\Q(t)^2dt^2-\Q^2(t)dr^2-R^2(t,r)(d\vartheta^2+\sin^2\vartheta d\varphi^2),
\ee
with $R(t,r)=\Q(t)f_{\rm k}(r)$. Here $f_{\rm k}=\{r,\sin(r),\sinh(r)\}$
for ${\rm k}=0,1,-1$, which corresponds, respectively, to spatially 
flat, closed, or open FRW universe.  The metric is invariant under 
spatial rotations and translations. 
We also assume the matter to be a homogeneous and isotropic perfect fluid,
so that 
\be
8\pi GT^{{\rm (m)}\rho}_{~~~~\lambda}={\rm diag }[\rho(t),-P(t),-P(t),-P(t)]. 
\ee
The conservation condition \eqref{matter0} then reduces to 
\be                            \label{matter}
\dot{\rho}+3\frac{\dot{\Q}}{\Q}(\rho+P)=0.
\ee
As for the reference  metric, we assume it to be only spherically symmetric,
but not necessarily homogeneous, so that 
\bea                                        \label{ff}
ds_f^2&=&dT^2(t,r)-dU^2(t,r)-U^2(t,r)(d\vartheta^2+\sin^2\vartheta d\varphi^2)= \\
&=&(\dot{T}^2-\dot{U}^2)dt^2+2(\dot{T}T^\prime-\dot{U}U^\prime)dtdr
+(T^{\prime 2}-U^{\prime 2})dr^2-U^2(t,r)(d\vartheta^2+\sin^2\vartheta d\varphi^2). \notag 
\eea
Therefore, the two metrics do not have the same Killing symmetries. 
The functions $T(t,r)$ and $U(t,r)$ are related to the St\"uckelberg fields,
$X^A=\{X^0,X^k\}=\{T(t,r),U(t,r)n^k\}$, where  $n^k=
(\sin\vartheta\cos\varphi,\cos\vartheta\sin\varphi,\cos\vartheta)$.

The above expressions imply that  
\be                                  \label{gamma2}
(\gamma^2)^\mu_{~\nu}=g^{\mu\sigma}f_{\sigma\nu}=\left(
\begin{array}{cccc}
A & C & 0 & 0 \\
-C & B & 0 & 0 \\
0 & 0 & {U^2}/{R^2} & 0 \\
0 & 0 & 0 & {U^2}/{R^2}
\end{array}
\right)\,,
\ee
where 
\be                            \label{ABC}
A=\frac{\dot{T}^2-\dot{U}^2}{\Q^2},~~~~~
C=\frac{\dot{T}T^\prime-\dot{U}U^\prime}{\Q^2},~~~~~
B=\frac{{U}^{\prime 2}-{T}^{\prime 2}}{\Q^2}.
\ee
To take the square root of this matrix, one makes the ansatz 
\be                                  \label{gamma}
\gamma^\mu_{~\nu}=\left(
\begin{array}{cccc}
a & c & 0 & 0 \\
-c & b & 0 & 0 \\
0 & 0 & u & 0 \\
0 & 0 & 0 & u
\end{array}
\right)\,.
\ee
Inserting this to \eqref{gamma2} gives algebraic equations 
\bea                     \label{ABC1}
a^2-c^2=A,\notag \\
c(a+b)=C,\notag \\
b^2-c^2=B,\notag \\
u^2=U^2/R^2,
\eea
whose solution is
\bea                                \label{abc}
a=\frac{A+\Delta}{Y},~~~~b=\frac{B+\Delta}{Y},~~~~~c=\frac{C}{Y},~~~~u=\frac{U}{R},
\eea
with
\be                                   \label{abc1}
Y=\sqrt{A+B+2\Delta},~~~~~ \Delta=\sqrt{AB+C^2}.
\ee

\section{Solution for the metric  \label{met}}
\setcounter{equation}{0}

Since $\gamma^\mu_{~\nu}$ is known, we can compute 
the energy-momentum tensor in \eqref{TT3}. 
It is convenient to lift one index and consider 
$T^\mu_\nu=g^{\mu\alpha}T_{\alpha\nu}$. 
To begin with, the eigenvalues of ${\cal K}^\mu_\nu=\delta^\mu_\nu-\gamma^\mu_{~\nu}$ are
\be
\lambda_{1,2}=\frac{2-a-b\pm\sqrt{(a-b)^2-4c^2}}{2},~~~\lambda_3=\lambda_4=u\,,
\ee
which can be complex-valued. However, the symmetric polynomials in 
\eqref{U2} are always real,  
\bea                      \label{LLL}
{\cal U}_2&=&u(u+2a+2b-6)+c^2+ab-3a-3b+6\,, \notag \\
{\cal U}_3&=&\alpha_3(1-u)[(a+b-2)u+2c^2+2ab-3a-3b+4]\,, \notag \\
{\cal U}_4&=&\alpha_4(1-u)^2(c^2+ab-a-b+1).
\eea
Inserting this and \eqref{gamma} into \eqref{TT3}
gives the following non-zero components:
\begin{align}      \label{tau}           
T^0_0&=u(6-2b-u)+3b-6+\alpha_3(u-1)(bu-2u-3b+4)+\alpha_4(u-1)^2(b-1),\notag \\
T^r_r&=u(6-2a-u)+3a-6+\alpha_3(u-1)(au-2u-3a+4)+\alpha_4(u-1)^2(a-1),\notag \\
T^0_r&=-c[(\alpha_3+\alpha_4)u^2-2(\alpha_4+2\alpha_3+1)u+3+3\alpha_3+\alpha_4]\,, \\
T^\vartheta_\vartheta&=T^\varphi_\varphi=
u(3-a-b)-6-ab-c^2+3a+3b+\alpha_4(u-1)(c^2+ab-a-b+1)\notag \\
&+\alpha_3[u(c^2+ab+3-2a-2b)-4-2ab+3b-2c^2+3a]. \notag  
\end{align}
It turns out that
\be
c(T^0_0-T^r_r)=(a-b)T^0_r\,.
\ee
The field equations 
$
G^\mu_\nu=m^2T^\mu_\nu+8\pi G T^{\rm (m)\mu}_{~~~~\nu}
$
imply that $T^\mu_\nu$ should be diagonal, since both $G^\mu_\nu$ and 
$T^{\rm (m)\mu}_{~~~~\nu}$ are diagonal. Therefore, one should have 
\be                       \label{TT0r}                                       
T^0_r=0,
\ee
and so $T_0^0=T^r_r$ if $c\neq 0$. 
Now, $T^0_r$ in \eqref{tau} will vanish if either 
$c=0$, or if the expression between
the brackets vanishes. The $c=0$ case will be considered in the Appendix.
If $c\neq 0$, then $T^0_r$ in \eqref{tau} will vanish if 
\be                                 \label{u0}
P_2(u)\equiv (\alpha_3+\alpha_4)u^2-2(\alpha_4+2\alpha_3+1)u+3+3\alpha_3+\alpha_4=0,
\ee
which requires that $u$ is constant, 
\be                                     \label{u}
u=\frac{1+2\alpha_3+\alpha_4\pm\sqrt{1+\alpha_3^2+\alpha_3-\alpha_4}}{\alpha_3+\alpha_4}\,,
\ee
which is real for 
$
\alpha_4\leq \alpha_3^2+\alpha_3+1. 
$
Inserting this to \eqref{tau} we find that the two components 
\be
T^0_0=T^r_r=(1-u)(u+u\alpha_3-\alpha_3-3)
\ee
are also constant. 
The conditions $\nabla_\mu T^\mu_\nu=0$ 
reduce in this case to 
\be                         \label{cons0}
{\nabla}_\mu T^{\mu}_{0}=2\frac{\dot{\Q}}{\Q}
\left(T^0_0-T^\vartheta_\vartheta\right)=0,
\ee
which requires that $T^0_0=T^\vartheta_\vartheta$. Now, using the above formulas
we find that 
\be                              \label{constr}
T^0_0-T^\vartheta_\vartheta=\frac{u+u\alpha_3-2-\alpha_3}{1-u}\,[(u-a)(u-b)+c^2 ],
\ee
and for this to be zero either the first or the second factor on the right 
must vanish. The first of these options implies a restriction on values 
of the parameters $\alpha_3,\alpha_4$. This will be discussed in the next Section. 
If we do not want the parameters to be restricted, then we should require that 
\be                       \label{uuu}
(u-a)(u-b)+c^2=0. 
\ee
This condition guarantees  that $T^\mu_\nu$ is conserved, so that this is the equation
for the  St\"uckelberg fields. 
Assuming that this condition is fulfilled, $T^\mu_\nu$ becomes proportional 
to the unit tensor and  
the field equations 
reduce to 
\begin{align}
G^\rho_\lambda&=\Lambda \delta^\rho_\lambda
+8\pi G T^{{\rm (m)}\,\rho}_{~~~~\lambda} \, \label{ee1} 
\end{align}
with 
\be                              \label{Lambda}
\Lambda=m^2(1-u)(u+u\alpha_3-\alpha_3-3).
\ee
The functions $a,b,c$ effectively drop out and only 
the constant $u$ remains. 
As a result, the effect of the graviton mass is the same as that of 
a constant cosmological term. Einstein equations \eqref{ee1} further reduce to 
the Friedmann equation 
\be                                    \label{FRW}
3\,\frac{\dot{\Q}^2+{\rm k}\Q^2}{\Q^4}=\Lambda+\rho,
\ee
where $\rho(\Q)$ is determined by the conservation condition \eqref{matter}.  
This equation describes a universe filled with matter and 
containing a
cosmological term mimicked by the graviton mass. 
At early times the matter density $\rho$ dominates, but 
in the long run the cosmological term 
wins, leading  to the self-acceleration. 

\section{Solution for the St\"uckelberg fields   \label{stueck}}
\setcounter{equation}{0}

The above solution for the metric is only possible if equation \eqref{uuu}
is satisfied, so that the whole procedure is consistent if only 
this equation can be fulfilled. 
Rewriting it as 
\be
ab+c^2+u^2=(a+b)u
\ee
and using expressions \eqref{abc} for $a,b,c$ yields 
\be                                \label{6.2}     
\frac{(A+\Delta)(B+\Delta)+C^2}{Y^2}+u^2=uY.
\ee
Now, using \eqref{abc1} one has 
\bea
(A+\Delta)(B+\Delta)+C^2&=&AB+C^2+(A+B)\Delta+\Delta^2\notag \\
&=&(A+B)\Delta+2\Delta^2=\Delta(A+B+2\Delta)=\Delta Y^2\,,
\eea
and so \eqref{6.2} becomes 
\be                   \label{eee}
\Delta+u^2=uY. 
\ee
Next, using \eqref{ABC}, one finds 
\be
AB+C^2=
\frac{(\dot{T}U^\prime-\dot{U}T^\prime)^2 }{\Q^4 },
\ee
and therefore 
\be
\Delta=\sqrt{AB+C^2}=\frac{\dot{T}U^\prime-\dot{U}T^\prime}{\Q^2 },
\ee
while 
\be
Y=\sqrt{A+B+2\Delta}=\frac{1}{\Q}\,\sqrt{ (\dot{T}+U^\prime)^2-(T^\prime+\dot{U})^2 },
\ee
so that \eqref{eee} reduces to 
\be                            \label{oops}
\dot{T}U^\prime-\dot{U}T^\prime+u^2\Q^2=u\Q\sqrt{ 
(\dot{T}+U^\prime+T^\prime+\dot{U})(\dot{T}+U^\prime-T^\prime-\dot{U}).
}
\ee
Squaring this finally gives 
\be                          \label{EEE}
(\dot{T}U^\prime-\dot{U}T^\prime+u^2\Q^2)^2-u^2\Q^2{ 
(\dot{T}+U^\prime+T^\prime+\dot{U})(\dot{T}+U^\prime-T^\prime-\dot{U})=0.
}
\ee
This is a quadratic PDE for $T(t,r)$, with the  coefficients 
$\Q(t)$ and $U=uR=u\Q(t)f_{\rm k}(r)$ determined by 
the solution of the Einstein equation \eqref{FRW}. 
Although this equation looks complicated, 
some of its solutions can be obtained. 

Let us first consider the case of spatially flat universe,
when $U=u\Q(t)r$. One makes the ansatz,
\be
T(t,r)=f(t)+{\cal C}\Q(t)r^2,
\ee
with constant ${\cal C}$.  In this case Eq.\eqref{EEE} reduces to 
\be
u^2\dot{\Q}^2-4{\cal C}\dot{\Q}\dot{f}+4{\cal C}^2\Q^2=0,
\ee
which can be resolved with respect to $f(t)$ to give 
\be                    \label{T1}
T(t,r)={\cal C}\int^t \frac{\Q^2}{\dot{\Q}}\,dt
+\left(\frac{u^2}{4{\cal C}}+{\cal C}r^2 \right)\Q\,.
\ee
This solution agrees with the one
obtained in \cite{gaba} for $\alpha_3=\alpha_4=0$, 
when $u=3/2$.

When the universe is spatially closed and
$U=u\Q(t)\sin(r)$, one assumes
\be
T(t,r)=f(t)+{\cal C}\Q(t)\cos(r),
\ee
which reduces Eq.\eqref{EEE} to 
\be
({\cal C}^2+u^2)(\dot{\Q}^2+\Q^2)=\dot{f}^2\,,
\ee
hence
\be                   \label{T2}
T(t,r)=\pm \int^t \sqrt{({\cal C}^2+u^2)(\dot{\Q}^2+\Q^2)}\, dt+{\cal C}\Q\cos(r).
\ee

When the universe is open and
$U=u\Q(t)\sinh(r)$, one makes the ansatz 
\be
T(t,r)=f(t)+{\cal C}\Q(t)\cosh(r),
\ee
reducing the problem to 
\be
({\cal C}^2-u^2)(\dot{\Q}^2-\Q^2)=\dot{f}^2,
\ee
so that 
\be                   \label{T3}
T(t,r)=\int^t \sqrt{({\cal C}^2-u^2)(\dot{\Q}^2-\Q^2)}\, dt+{\cal C}\Q\cosh(r).
\ee
This completes our construction, since we have determined the metric
and the St\"uckelberg fields for all three types of the universe and for generic
$\alpha_3,\alpha_4$. 

All above considerations require that 
$
\alpha_4\leq \alpha_3^2+\alpha_3+1
$, 
since otherwise $u$ in \eqref{u} becomes complex-valued. 
Let us consider the limit where this inequality is saturated, 
$\alpha_4=\alpha_3^2+\alpha_3+1$. This implies that $u+u\alpha_3-2-\alpha_3=0$,
in which case the conservation condition \eqref{cons0}
will be fulfilled without imposing the constraint \eqref{uuu},
because the first factor in \eqref{constr} will then vanish. 
This possibility has been first analyzed  
in \cite{cosm} (and recently rediscovered in \cite{Kobayashi}). 
The obtained 
above general solution applies in this case too, and the 
physical metric is determined by the same equations \eqref{FRW}
as before. However, the function $T(t,r)$ needs not to be now 
the same as before, since 
the constraint \eqref{uuu} is no longer imposed, so that 
there is actually no condition for 
$T(t,r)$. Therefore, $T(t,r)$ remains
arbitrary
(a particular choice $T=-\int \dot{U} dr$ was made in 
\cite{cosm}).

 \section{Solution in the tetrad formulation}
\setcounter{equation}{0}

Let us now see how the same results are obtained within the tetrad formulation 
used in \cite{vvv}. This formulation was originally introduced in \cite{CM}
and then further  developed in \cite{tetrad}, 
it is equivalent to the standard formulation but can sometimes be more efficient 
\cite{cosm},\cite{cosm1}.  Although originally it was used to argue 
in favor of a possible presence of the ghost \cite{CM},
the absence of ghost in the tetrad description was shown in  \cite{tetrad}.

The basic variables in the tetrad formulation  
are two tetrads  $e_A^\mu$ and $\omega^B_\nu$ 
which determine the two metrics,
\be
g^{\mu\nu}=\eta^{AB}e_A^\mu e_B^\nu,~~~~~~~
f_{\mu\nu}=\eta_{AB}\omega^A_\mu \omega^B_\nu,
\ee 
and also the tensor 
\be                                 \label{gamtil}
\tilde{\gamma}^\mu_{~\nu}=e^\mu_A\omega^A_\nu\,.
\ee
Defining ${\cal K}^\mu_\nu=\delta^\mu_\nu-\tilde{\gamma}^\mu_{~\nu}$,
the action is still given by Eq.\eqref{action}, with the interaction 
\be
{\cal U}=\frac{1}{2}\,(({\cal K}^\mu_\mu)^2-{\cal K}^\mu_\nu{\cal K}^\nu_\mu)
-\frac{\alpha_3}{3!}\,\epsilon_{\mu\nu\rho\sigma}
\epsilon^{\alpha\beta\gamma\sigma}
{\cal K}_{\alpha}^{\mu}{\cal K}_{\beta}^{\nu}{\cal K}_{\gamma}^{\rho}
-\frac{\alpha_4}{4!}\,\epsilon_{\mu\nu\rho\sigma}
\epsilon^{\alpha\beta\gamma\delta}{\cal K}_{\alpha}^{\mu}
{\cal K}_{\beta}^{\nu}{\cal K}_{\gamma}^{\rho}{\cal K}_{\delta}^{\sigma}\,
\ee
(the notation $c_3=-\alpha_3$, $c_4=-\alpha_4$ was used in \cite{vvv}). 
Comparing with \eqref{int},\eqref{U3}, this is the same expression as before, 
up to the replacement $\gamma^\mu_{~\nu}\to\tilde{\gamma}^\mu_{~\nu}$. 
Varying the action with respect to $e_A^\mu$ is straightforward and 
gives the Einstein equations 
$
G^\mu_\nu=m^2T^\mu_\nu+8\pi G T^{\rm (m)\mu}_{~~~~\nu}\,
$
with
\be                                      \label{tau3}
T^\mu_\nu=e^\mu_A\frac{\delta {\cal U}}{\delta e^\nu_A}-{\cal U}\,\delta^\mu_\nu\,.
\ee
Explicitly, 
\be                               \label{T34}
T_{\mu\nu}=
\tilde{\gamma}_{\mu\alpha}\left\{{\cal K}^\alpha_\nu-[{\cal K}]\delta^\alpha_\nu
+\alpha_3 {\Xi}^\alpha_\nu+\alpha_4 {\Omega}^\alpha_\nu
\right\}\,-{\cal U}\,g_{\mu\nu}\,,
\ee
where $\tilde{\gamma}_{\mu\alpha}=g_{\mu\nu}\tilde{\gamma}^\nu_{~\alpha}$ and  
${\Xi}^\alpha_\nu$ and ${\Omega}^\alpha_\nu$ are defined by \eqref{tau2}. 
This expression agrees with that 
in \eqref{tau1}, 
up to the replacement $\gamma^\mu_{~\nu}\to\tilde{\gamma}^\mu_{~\nu}$. 
Therefore, if we could show that $\tilde{\gamma}^\mu_{~\nu}=\gamma^\mu_{~\nu}$,
this would mean that we have the same equations as before. 

The equality $\tilde{\gamma}^\mu_{~\nu}=\gamma^\mu_{~\nu}$
can be enforced  by the 
the following condition:
\be                            \label{symmetry}
\omega_{A\mu}e^\mu_B=\omega_{B\mu}e^\mu_A\,,
\ee
with $\omega_{C\mu}=\eta_{CA}\omega^A_\mu$, because 
\be
\tilde{\gamma}^\mu_{~\alpha}\tilde{\gamma}^\alpha_{~\nu}=
e^\mu_A\omega^A_\alpha e^\alpha_B\omega^B_\nu=
e^{A\mu}\omega_{A\alpha} e^\alpha_B\omega^B_\nu=
e^{A\mu}\omega_{B\alpha} e^\alpha_A\omega^B_\nu=g^{\mu\alpha}f_{\alpha\nu}\,,
\ee
and therefore   $\tilde{\gamma}^\mu_{~\nu}$ fulfills the very same equation  which defines  
${\gamma}^\mu_{~\nu}$. The condition \eqref{symmetry} was
originally postulated in \cite{CM}. However, 
it turns out that it actually follows from the field equations 
(see also \cite{tetrad}). Indeed, the Einstein equations imply that $T_{\mu\nu}$ is symmetric.  
The expression in \eqref{T34} will be always symmetric if only the first four
terms on the right are separately symmetric.  
Therefore, $\tilde{\gamma}_{\mu\nu}$ should be symmetric, and this guarantees 
that the other three terms are symmetric as well. 
As a result, the field equations require that 
\be
g_{\mu\alpha}\tilde{\gamma}^\alpha_{~\nu}=g_{\nu\alpha}\tilde{\gamma}^\alpha_{~\mu}\,.
\ee
Using the definition of $\tilde{\gamma}^\alpha_{~\nu}$ and also 
$g_{\mu\nu}=\eta_{AB}e_\mu^A e_\nu^B$ where $e_\mu^A$ is the inverse of $e^\mu_A$,
these relations assume the form  
\be
e^C_\mu\omega_{C\nu}=e^C_\nu\omega_{C\mu}\,,
\ee
multiplying which by $e^\mu_Ae^\nu_B$ 
gives precisely Eq.\eqref{symmetry}. 
 Therefore, the equality $\tilde{\gamma}^\mu_{~\nu}=\gamma^\mu_{~\nu}$
is imposed dynamically, hence
we can remove the tilde sign and conclude that  
the tetrad formulation 
gives the same theory as before. 

To obtain the solution, one makes the ansatz, 
\bea               
e_0&=&\frac{1}{\Q}\,\frac{\partial}{\partial t },~~~
e_1=\frac{1}{\Q}\,\frac{\partial}{\partial r },~~~
e_2=\frac{1}{R}\,\frac{\partial}{\partial \vartheta },~~~
e_3=\frac{1}{R\sin\vartheta}\,\frac{\partial}{\partial \varphi },~ \\
\omega^0&=&\Q\,(a\,dt+c\,dr),~~~~\omega^1=\Q\,(b\,dr-c\,dt), ~~~~~
\omega^2=uR\,d\vartheta,
~~~~\omega^3=uR\sin\vartheta\, d\varphi\,,    \notag                \label{tetrads}
\eea 
which fulfills \eqref{symmetry}, 
with $\Q=\Q(t)$, and where $R=\Q(t)f_{\rm k}(r)$ is the same as in Eq.\eqref{gg}. This implies that 
$g_{\mu\nu}$ is the same as in \eqref{gg}. Computing $\gamma^\mu_{~\nu}=e^\mu_A\omega^A_\nu$
then gives the same result as in \eqref{gamma}, and therefore all analysis of 
Section \ref{met} goes through without any changes. 

There remains to analyze the consistency condition \eqref{uuu}. Let us remember 
that the metric $f_{\mu\nu}=\eta_{AB}\omega^A_\mu \omega^B_\nu$ should be flat. 
However, this does not mean that 1-forms $\omega^A=\omega^A_\mu dx^\mu$
coincide with differentials of the  St\"uckelberg scalars, $dX^A$, because 
it is still possible to perform local Lorentz rotations, so that one actually has
$\omega^A=L^A_{~B}dX^B$, where $L^A_{~B}$ is a position-dependent 
$SO(1,3)$ matrix. Comparing with $f_{\mu\nu}$ in \eqref{ff},
it follows that $\omega^2=Ud\vartheta$ and $\omega^3=U\sin\vartheta d\varphi$,
while $\omega^0$ and $\omega^1$ can differ from $dT$ and $dU$ by a local Lorentz boost,
\be
\omega^0=\cosh(\alpha) dT+\sinh(\alpha)dU,~~~~~~\omega^1=\cosh(\alpha) dU+\sinh(\alpha)dT\,,
\ee
where $\alpha$ is the boost parameter. Explicitly, 
\bea                                   \label{end}
\Q\,(a\,dt+c\,dr)&=&\cosh(\alpha)(\dot{T}dt+T^\prime dr)+\sinh(\alpha)(\dot{U}dt+U^\prime dr),\notag \\
\Q\,(b\,dr-c\,dt)&=&\cosh(\alpha)(\dot{U}dt+U^\prime dr)+\sinh(\alpha)(\dot{T}dt+T^\prime dr). 
\eea
Comparing the coefficients in front of $dt,dr$ gives four conditions, which determine 
$a,b,c,\alpha$ in terms of $T,U$. Inserting the resulting $a,b,c$ 
to \eqref{uuu} 
then gives precisely the same equation for $T(t,r)$ as in \eqref{oops}. 
Therefore, we recover the same results as before. 

In the case when $\alpha_4=\alpha_3^2+\alpha_3+1$ discussed at the end of Section \ref{stueck},
when the constraint \eqref{uuu} is not imposed, 
there is no condition 
for coefficients $a,b,c,\alpha$ obtained from \eqref{end}, so that the choice of 
$T(t,r)$ remains arbitrary. 
Equivalently, one can choose some value of $\alpha$ and then calculate $T(t,r)$ from \eqref{end}. 
For example, setting $\alpha=0$ requires that $T^\prime+\dot{U}=0$ 
and so  $T=-\int \dot{U} dr$ \cite{cosm}.  

\section{Conclusion}

We presented the most general cosmological solution
in the dRGT ghost-free massive gravity 
for which the physical metric is homogeneous and isotropic 
but the St\"uckelberg fields are inhomogeneous.    
The solution includes the matter source, it exists for generic values of the 
theory parameters,  
and its physical metric describes a FRW universe 
that can be spatially flat, open or closed, and which shows 
the late-time acceleration due to the effective cosmological term mimicked
by the graviton mass. 

Even though the physical metric is homogeneous and isotropic, 
its perturbations are expected to be 
inhomogeneous, due to the inhomogeneous structure of the 
St\"uckelberg fields \cite{gaba}. 
This effect will be proportional to $m^2$ and so will be small in small enough 
regions of space. However, it would be still
interesting to compute the linear perturbation spectrum (see also \cite{amico}). 
It would also be interesting to see if the above construction could be generalized 
to describe non-linear anisotropic deformations of the homogeneous and 
isotropic background. 

We have also shown the equivalence between the standard parametrization
of the dRGT theory and the tetrad approach (see also \cite{tetrad}). 

\renewcommand{\thesection}{APPENDIX. SOLUTIONS WITH TWO DIAGONAL METRICS}
\section{}
\renewcommand{\theequation}{A.\arabic{equation}}
\setcounter{equation}{0}
\setcounter{subsection}{0}

For the sake of completeness, 
we review in this Appendix solutions for which 
 both the physical and reference  metrics are simultaneously diagonal, homogeneous 
and isotropic. Most of them  have been previously reported in the literature,
although solutions of type II described below are new.
All of them 
are obtained by setting $c=0$ in formulas 
of Sections \ref{sym},\ref{met}. 
Such solutions exist only for particular values of the 
spatial curvature k and/or show a degenerate 
reference  metric. Therefore, they are less interesting 
physically than solutions  described in the main text above.

If $c=0$, then the condition $T^0_r=0$ in \eqref{TT0r}
will be fulfilled without imposing the constraint \eqref{u0}, while  
Eqs.\eqref{ABC1} will imply  that $C=0$, so that 
the metric $f_{\mu\nu}$ in \eqref{ff} is diagonal.  
Eqs.\eqref{ABC},\eqref{ABC1},\eqref{abc},\eqref{abc1} then reduce to 
\begin{subequations}                \label{A1}
\begin{align}
&a^2=\frac{\dot{T}^2-\dot{U}^2}{\Q^2},~~~~~~~~\label{A1a}\\
&b^2=\frac{{U}^{\prime 2}-{T}^{\prime 2}}{\Q^2},~~~~~~\label{A1b}\\
&\dot{T}T^\prime=\dot{U}U^\prime\,,               \label{A1c}
\end{align}
\end{subequations}
with $U=u\Q(t)f_{\rm k}(r)$. 
Here $a,b,u$ are functions of $t,r$. 
The energy-momentum tensor is defined by Eqs.\eqref{tau} with $c=0$
and should satisfy the 
 isotropy condition,
\be                         \label{A2}
T^r_r=T^\vartheta_\vartheta,
\ee
the homogeneity condition, $(T^\mu_\nu)^\prime=0$, and the 
conservation condition,
\be                          \label{A3}
\dot{T}^0_0+3\,\frac{\dot{\Q}}{\Q}\,(T^0_0-T^r_r)=0.
\ee
The Einstein equations for the metric $g_{\mu\nu}$ then 
reduce to 
\be                                    \label{A4}
3\,\frac{\dot{\Q}^2+{\rm k}\Q^2}{\Q^4}=m^2T^0_0
+\rho.
\ee
Equations \eqref{A1}--\eqref{A4} should be solved to  determine $a,b,u$, $\Q(t)$, 
and $T(t,r)$. 

Let us consider first  the isotropy condition \eqref{A2}. 
Using Eqs.\eqref{tau} gives
\be                          \label{App0}
T^r_r-T^\vartheta_\vartheta=(u-b)
[(au-2a-2u+3)\alpha_3+(1-u-a+au)\alpha_4+3-a-u],
\ee
and since this has to be zero, 
either the first or second factor on the right
should vanish, which we shall call case I and case II, respectively. 

In the 
{case I} one has  $u=b$ and the conservation 
condition \eqref{A3} reduces to 
\be                           \label{App1} 
(\dot{\beta}-\dot{\Q}a )P_2(u)
=0,
\ee
with $\beta=u\Q$ and $P_2(u)$ being defined by Eq.\eqref{u0}. 
 Depending on which of the two factors on the left vanishes, 
there are two subcases to analyze, 
let us call them Ia and Ib. 

\noindent
\underline{\bf Case Ia:} $\dot{\beta}-\dot{\Q}a=0$. One has
$a=\dot{\beta}/\dot{\Q}$ and $b=u=\beta/\Q$. 
Inserting this to \eqref{A1a} and \eqref{A1b} gives 
\be
{T}^\prime=\beta\sqrt{f_{\rm k}^{\prime 2}-1 },~~~~~~~~~
\dot{T}=\dot{\beta}\sqrt{  {\Q^2}/{\dot{\Q}^2 }+f_{\rm k}^2 }.
\ee
These should fulfill \eqref{A1c} and also the integrability conditions 
$\partial^2_{tr}T=\partial^2_{rt}T$, 
which is only possible if  
$\dot{\beta}=0$. Therefore, the St\"uckelberg scalars are
\be
T(r)=\beta\int dr\sqrt{f_{\rm k}^{\prime 2}-1 },
~~~~~~~~U(r)=\beta f_{\rm k}\,,
\ee
with constant $\beta$. The scale factor satisfies \eqref{A4}
with 
\be                                    \label{FRW1}
T^0_0=-4\alpha_3-\alpha_4-6
+\frac{3\beta(3\alpha_3+\alpha_4+3)}{\Q}
-\frac{3\beta^2(1+\alpha_4+2\alpha_3)}{\Q^2}
+\frac{\beta^3(\alpha_3+\alpha_4)}{\Q^3}\,,
\ee
and for ${\rm k}=0,\pm 1$. 
This solution was found in \cite{cosm} (with the notation $c_3=\alpha_3$,
$c_4=-\alpha_4$). 
Its stability has been studied in 
\cite{cosmstab}. 
Although its physical metric is well behaved, 
the reference  metric is degenerate, since both $T$ and $U$ 
do not depend on $t$. Moreover, $T$ becomes imaginary for ${\rm k}=1$.

 \noindent
\underline{\bf Case Ib:} $P_2(u)=0$. 
This gives the same equation for $u$ as in \eqref{u0}.
Therefore, one has $b=u$ 
given by \eqref{u}. Eqs.\eqref{A1b},\eqref{A1c} then yield 
\be
T^\prime =u\Q\sqrt{f_{\rm k}^{\prime 2}-1 },~~~~~~~~
\dot{T}=\frac{u\dot{\Q}f_{\rm k}f_{\rm k}^\prime}
{\sqrt{f_{\rm k}^{\prime 2}-1 }},
\ee
and the integrability condition $\partial^2_{tr}T=\partial^2_{rt}T$
reduces to 
$
\left(\sqrt{f_{\rm k}^{\prime 2}-1 }\right)^{\prime\prime}
=\sqrt{f_{\rm k}^{\prime 2}-1 },
$
which can be fulfilled only for ${\rm k}=-1$, when $f_{\rm k}=\sinh(r)$. 
As a result, the St\"uckelberg scalars are
\be
T=u\,\Q(t)\cosh(r),~~~~~~~U=u\,\Q(t)\sinh(r).
\ee
The scale factor fulfills \eqref{A4} with ${\rm k}=-1$ and 
with $m^2T^0_0=\Lambda$, 
where $\Lambda$ is the same as in \eqref{Lambda}. 
This solution was found in 
\cite{Muk}. The reference  metrics is non-degenerate, 
but the solution exists only for ${\rm k}=-1$. Moreover, it
shows a non-linear instability \cite{Muk}. 

Let us now consider the case II, when the second factor in \eqref{App0} vanishes,
and so 
\be                              \label{A12}
a=\frac{(1+\alpha_4+2\alpha_3)u-3-\alpha_4-3\alpha_3 }
{(\alpha_3+\alpha_4)u-1-\alpha_4-2\alpha_3 }\,\equiv a(u).
\ee
The conservation condition \eqref{A3} 
then reduces to 
\be
(b-a)\left(\dot{u}-u_1\right)P_1(u)=0\,,
\ee
where 
\be
u_1=-3\,\frac{\dot{\Q}}{\Q}\,\frac{P_2(u)}{P_1(u)},~~~~~~~~~
P_1(u)=2(\alpha_3+\alpha_4)u-4\alpha_3-2\alpha_4-2
=\frac{dP_2(u)}{du}.
\ee
Therefore, there are two options, either $a=b$ or $\dot{u}=u_1$,
since one cannot have $P_1(u)=0$, because $a$ diverges in this case. 

 \noindent
\underline{\bf Case IIa:} $b=a$. In this case Eqs.\eqref{A1} 
yield $T(t,r)=U(t,r)$, so that that the reference metric 
 is degenerate. 
This also implies that $a=b=0$, in which case one obtains 
from \eqref{A12} $u=(3\alpha_3+\alpha_4+3)/(1+2\alpha_3+\alpha_4)$. 
The scale factor fulfills Eq.\eqref{A4} with 
\be
T^0_0=\frac{\alpha_3^2+2\alpha_3-\alpha_4+3}{2\alpha_3+\alpha_4+1}.
\ee
This solution is new and has not been reported in the literature before. 

 \noindent
\underline{\bf Case IIb:} $\dot{u}=u_1$. Integrating this equation gives
the relation between $u$ and $\Q$, 
\be                             \label{A16}
P_2(u)={{\cal A}}/{\Q^3}\,,
\ee
where ${\cal A}$ is the integration constant, so that $u=u(t)$
and therefore $a=a(t)$. Eqs.\eqref{A1a} and \eqref{A1c} then yield 
\be
\dot{T}=\sqrt{\alpha^2+\dot{\beta}^2f_{\rm k}^2},~~~~~~~~~~
T^\prime=
\frac{\beta\dot{\beta}f_{\rm k}f_{\rm k}^\prime}
{\sqrt{\alpha^2+\dot{\beta}^2f_{\rm k}^2}}\,,
\ee
with $\alpha(t)=a\Q$ and $\beta(t)=u\Q$.
The integrability condition $\partial^2_{tr}T=\partial^2_{rt}T$
reduce to 
$
\dot{\beta}={\cal B}\alpha\,,
$
where ${\cal B}$ is an integration constant. Using definitions
of $\alpha,\beta$, this transforms to 
\be                              \label{A18}
\dot{\Q}=\frac{{\cal B}\Q \,a(u)P_1(u)}{uP_1(u)-3P_2(u)}\,,
\ee 
while Eq.\eqref{A1b} yields
\be
b=\frac{uf_{\rm k}^\prime}{\sqrt{1+{\cal B}^2f_{\rm k}^2}}.
\ee
Since $b$ depends both on $t$ and $r$, 
there are three options to consider. The first two 
correspond 
to choosing either ${\rm k}={\cal B}=0$,
or ${\rm k}=-1$ and ${\cal B}^2=1$, in which cases $b=u(t)$. 
The last option is to let 
$b$ depend on $r$, but to set  
$P_2(u)=0$, in which case the coefficient in front of $b$ in 
$T^0_0$ vanishes. In each of these three cases, using 
\eqref{A18} to eliminate $\dot{\Q}$ in the 
Einstein equation \eqref{A4} and taking into account
\eqref{A16}, one obtains an algebraic relation containing $\Q$
and the matter density $\rho$.  As a result, the solution exists 
only for fine-tuned $\rho$, which is unlikely to be physically 
interesting, so that we do not analyze this case any further.

\end{document}